\documentclass[%
twocolumn,
superscriptaddress,
nofootinbib,
 amsmath,amssymb,
 aps, prd
]{revtex4-2}

\usepackage{graphicx}
\usepackage{tensor}
\usepackage{siunitx}
\usepackage{breqn}
\DeclareUnicodeCharacter{2212}{-}
\usepackage{xcolor} % Remove after notes are no longer needed
\usepackage[normalem]{ulem} % Remove after notes are no longer needed
\usepackage{lipsum} % Remove after filler text is no longer needed
\usepackage{amsmath}
\usepackage{soul}
\usepackage[pdfborder={0 0 0}, plainpages=false, hyperfigures=true]{hyperref}

\newcommand{\beq}{\begin{equation}}
\newcommand{\eeq}{\end{equation}}
\newcommand{\beqn}{\begin{eqnarray}}
\newcommand{\eeqn}{\end{eqnarray}}

\newcommand{\lo}{\mathrel{\raise.3ex\hbox{$<$}\mkern-14mu
    \lower0.6ex\hbox{$\sim$}}}
\newcommand{\go}{\mathrel{\raise.3ex\hbox{$>$}\mkern-14mu
    \lower0.6ex\hbox{$\sim$}}}

\newcommand{\UNH}{\affiliation{Department of Physics, University of New Hampshire, 9 Library Way, Durham NH 03824, USA}}

\newcommand{\GRAPPA}{\affiliation{GRAPPA Institute of High-Energy Physics, University of Amsterdam, Science Park 904, \\ 1098 XH Amsterdam, Netherlands}}

\begin{document}
\title{A study of the agreement between binary neutron star ejecta models derived from numerical relativity simulations}
\author{Amelia Henkel} \UNH
\author{Francois Foucart}\UNH
\author{Geert Raaijmakers}\GRAPPA
\author{Samaya Nissanke}\GRAPPA

\begin{abstract}
Neutron star mergers have recently become a tool to study extreme gravity, nucleosynthesis, and the chemical composition of the Universe.  To date, there has been one joint gravitational and electromagnetic observation of a binary neutron star merger, GW170817, as well as a solely gravitational observation, GW190425.  In order to accurately identify and interpret electromagnetic signals of neutron star mergers, better models of the matter outflows generated by these mergers are required.  We compare a series of ejecta models to see where they provide strong constraints on the amount of ejected mass expected from a system, and where systematic uncertainties in current models prevent us from reliably extracting information from observed events.  We also examine 2396 neutron star equations of state compatible with GW170817 to see whether a given ejecta mass could be reasonably produced with a neutron star of said equation of state, and whether different ejecta models provide consistent predictions.  We find that the difference between models is often comparable to or larger than the error generally assumed for these models, implying better constraints on the models are needed. We also note that the extrapolation of outflow models outside of their calibration window, while commonly needed to analyze gravitational wave events, is extremely unreliable and occasionally leads to completely unphysical results.

\end{abstract}

\maketitle

\section{Introduction}
The gravitational observation of binary neutron star merger GW170817 and its corresponding kilonova signal AT2017gfo ushered the world into the era of multi-messenger astronomy (Abbott \textit{et al.} 2017 ~\cite{Abbott_2017}).  Binary neutron star (BNS) mergers in general are interesting for the wealth of information encoded in their signal.  The electromagnetic signal of BNS mergers can elucidate several source parameters, such as the mass of the outflows from the system, its composition, morphology, and potentially information about the fate of the system. 
Their gravitational wave signal can provide information about the masses of the component objects, their mass ratio, spins, and the fate of the remnant system.  In general, binary neutron star mergers contain rich physics, providing insight into the strong gravity regime; nucleosynthesis and the origin of heavy elements; the dynamics and formation mechanisms of gamma ray bursts (GRBs); and the neutron star equation of state (EOS).  

The ejecta and observed kilonova signal is highly dependent on the mass, velocity, morphology, and composition of the outflows (Wollaeger \textit{et al.} 2021 ~\cite{wollaeger2021broad}, Barnes and Kasen 2013 ~\cite{Barnes_2013}). We often model the merger ejecta as consisting of two components: the dynamical ejecta, ejected within milliseconds of the merger, and the disk wind ejecta, produced up to a few seconds after the merger.  The dynamical ejecta is typically faster, and may include a neutron-rich component associated with the tidal disruption of a neutron star and a less neutron-rich component due to shocks and oscillations in the forming neutron star remnant. The wind ejecta is typically slower, though both its velocities and composition may be significantly impacted by physical processes that are not fully included in many existing post-merger simulations (magnetohydrodynamics, neutrino transport)~\cite{Kasen_2017,Bulla_2019,Shibata:2021xmo}.  

Given the limited number of neutron star merger simulations performed so far, candidate electromagnetic signals to neutron star mergers are typically analyzed using approximate analytical model fitted to the result of numerical simulations. To date, a number of models have been developed to constrain the dynamics and outflows of binary neutron star mergers (e.g. ~\cite{Yamamoto_2008,Dietrich_2017a,Radice_2018,Kr_ger_2020,wollaeger2021broad,Nedora_2021}).
Recent work by Nedora \textit{et al.} 2021 ~\cite{Nedora_2021} clearly demonstrated that analytical fits attempting to predict the properties of neutron star merger ejecta can be heavily influenced by the level of microphysics included in the simulations used to derive that fit. In this manuscript, we attempt to answer a slightly different question, namely how robust inferences made about the properties of neutron stars are to the choice of fitting formula. This remains an important open question because some models do not include reliable error bars for their predictions, while models that do include errors bars only cover a limited region of parameter space and may lead to significantly larger systematic errors when used outside of their intended region of validity -- something that should ideally be avoided, yet is common practice when analyzing existing EM signals from neutron star mergers due to the lack of models covering the entire parameter space of interest.

We structure our paper in the following way: first, we introduce the ejecta models considered in this study.  We then present a comparison of these models, and show where they currently agree and disagree, under the simplifying assumption that both neutron stars have the same radius.  We later drop the constant radius assumption and apply the ejecta models to a series of equations of state consistent with gravitational wave observations of GW170817 to see if current observations can elucidate the true dense matter EOS.  Finally, we conclude with a discussion of our results, drawing attention to the cases most similar to BNS events GW170817 and GW190425.

\section{Ejecta models}

\begin{table*}[t]
\centering
\begin{tabular}{|c|c|c|c|c|c|}
\hline
    Source & Model & Simulations & Inputs & Calibrated Q & Uncertainty  \\
\hline
    DU~\cite{Dietrich_2017a} & Dyn. Ej. & 172 sims from ~\cite{Hotokezaka_2013,Bauswein_2013,Dietrich_2015,Lehner_2016,Sekiguchi_2016,Dietrich_2017b} & $M_1,M_2,C_1,C_2$ & 0.48-1.0 & $0.004M_\odot$ \\
    KF~\cite{Kr_ger_2020} & Dyn. Ej. & 200 sims from ~\cite{Dietrich_2017a,Kiuchi_2019} & $M_1,M_2,C_1,C_2$ & 0.48-1.0 & $0.004M_\odot$ \\
    NAL~\cite{Nedora_2021} & Dyn. Ej. & 324 sims from ~\cite{Nedora_2021,Sekiguchi_2015,Sekiguchi_2016,Vincent_2020,Radice_2018,Perego_2019,Bernuzzi_2020,Nedora_2019} & $\tilde{\Lambda},\tilde{Q}$ & 0.54-1.0 & $0.5M_{ej}+5\times10^{-5}M_\odot$ \\
    KF~\cite{Kr_ger_2020} & Disk & 57 sims from ~\cite{Kiuchi_2019,Radice_2018} & $M_{low},C_{low}$ & 0.775-1.0 & $0.5M_{disk}+5\times10^{-4}M_\odot$ \\
    DAL~\cite{Dietrich_2020} & Disk & 73 sims from ~\cite{Kiuchi_2019,Radice_2018,Dietrich_2017b,Hotokezaka_2011} & $Q,M_{tot},M_{th}$ & 0.571-1.0 & Not quoted \\
    RAL~\cite{Radice_2018} & Disk & 59 sims performed by RAL & $\tilde{\Lambda}$ & 0.85-1.0 &  $0.5M_{disk}+5\times10^{-4}M_\odot$ \\ 
    NAL~\cite{Nedora_2021} & Disk & 119 sims from ~\cite{Nedora_2021,Sekiguchi_2015,Sekiguchi_2016,Vincent_2020,Radice_2018,Perego_2019,Bernuzzi_2020,Nedora_2019} & $\tilde{\Lambda},\tilde{Q}$ & 0.54-1.0 & $0.5M_{disk}+(5\times10^{-4})M_\odot$ \\
\hline
\end{tabular}
\caption{Information about the dynamical ejecta and disk mass models considered in this work.  From left to right, the columns indicate (i) the name of the model; (ii) type of outflow; (iii) information about the simulations to which the models are calibrated; (iv) the models' input parameters; (v) the range of mass ratios to which the models are calibrated; and (vi) the uncertainty associated with each model, when quoted.}
\end{table*}

\subsection{Dynamical Ejecta Models}

First, we visit three fitting formulae for the amount of dynamical ejecta emitted from neutron star mergers, taken from Kruger and Foucart 2020 ~\cite{Kr_ger_2020} (hereafter KF), Dietrich and Ujevic ~\cite{Dietrich_2017a} (hereafter DU), and Nedora \textit{et al.} ~\cite{Nedora_2021} (hereafter NAL).  
We note that these formulae do not distinguish between the cold, neutron-rich tidal ejecta and the hot, less neutron-rich ejecta produced by the collision of the neutron star cores, and do not tell us anything about the geometry of the ejecta. We do not investigate here the impact of these issues on kilonovae light curves, but it is worth noting that even if these formulae were exact, they would not be sufficient to predict the kilonova signals associated with the dynamical ejecta. NAL introduced some fitting formulae providing partial information about the composition and geometry of the outflows, but they point out that the accuracy of these formulae is significantly limited by the small number of numerical relativity simulations with sufficiently advanced microphysics that reported the observables fitted in their work.

DU present a phenomenological fit of the mass, energy, and velocities of dynamical ejecta derived from a large series of numerical relativity simulations (N=172).  They use simulations presented in Hotokezaka \textit{et al.} 2013 ~\cite{Hotokezaka_2013}, Bauswein \textit{et al.} 2015 ~\cite{Bauswein_2013}, Dietrich \textit{et al.} 2015 ~\cite{Dietrich_2015}, Lehner \textit{et al.} 2016 ~\cite{Lehner_2016}, Sekiguchi \textit{et al.} 2016 ~\cite{Sekiguchi_2016}, and Dietrich \textit{et al.} 2017b ~\cite{Dietrich_2017b}.  Their data set combines results from general relativistic grid structured codes with results employing a smoothed particle hydrodynamics (SPH) code under the conformal flatness approximation.  It includes simplified EOSs and tabulated EOSs, as well as simulations with and without neutrino treatment.  As this study only considers dynamical ejecta, and no wind ejecta, Dietrich and Ujevic note that their estimates can provide a lower bound for the luminosity of EM observables.

Their fit for the dynamical ejecta mass is as follows: 
\begin{multline}
\frac{M_{ej}^{fit}}{10^{-3}M_\odot}=\left[a\left(\frac{M_2}{M_1}\right)^{1/3}\left(\frac{1-2C_1}{C_1}\right) \right.\\ \left. +b\left(\frac{M_2}{M_1}\right)^n+c\left(1-\frac{M_1}{M_1^*}\right)\right]+1\leftrightarrow 2 +d 
\label{eq:DUdynej}
\end{multline}
where $a=-1.35695$, $b=6.11252$, $c=−49.43355$, $d=16.1144$, and $n=−2.5484$; $M^*$ is the baryonic mass, M is the gravitational (ADM) mass in isolation, $C=GM/Rc^2$ is the compactness and $R$ is the radius of the neutron star. \\
KF developed formulae for the dynamical ejecta of BNS and black hole-neutron star (BHNS) systems, as well as the disk mass for BNSs.  Their dynamical ejecta fit for BHNS systems is outside the scope of this work, as we only consider BNSs; their disk mass ejecta fit is described in greater detail in Section 2.2.  KF's fit is similar to DU's, but does not depend on baryonic mass.  In their fitting formulae, they aim for analytical simplicity as well as physically reasonable extrapolation towards high-compactness stars. They also derive their fits from a slightly larger region of parameter space; namely, they base their work off of NR simulations from Radice \textit{et al.} 2018 ~\cite{Radice_2018} and Kiuchi \textit{et al.} 2019 ~\cite{Kiuchi_2019}, in addition to those used by DU. These additional simulations include, among other things, more asymmetric binaries than in the original data set.

Their dynamical ejecta fit for binary neutron stars is given as
\begin{multline}
    \frac{M_{dyn}}{10^{-3}M_\odot}=\left(\frac{a}{C_1}+b\frac{M_2^n}{M_1^n}+cC_1\right)M_1+(1\leftrightarrow 2)
    \label{eq:KFdynej}
\end{multline}
Here, a=−9.3335, b=114.17, c=−337.56, and n=1.5465.   

NAL presents fitting formulae for the dynamical ejecta of BNS mergers based on a polynomial in mass ratio $\tilde Q$ and reduced tidal deformability $\tilde{\Lambda}$, defined as 
\begin{equation}
    \tilde{\Lambda} = \frac{16}{13}\frac{(M_A+12M_B)M_A^4\Lambda_A}{M^5}+(A \leftrightarrow B). 
\end{equation}

In this convention, $\tilde Q=M_A/M_B \geq 1$ and $M=M_A+M_B$.  They obtain their fitting formula from a suite of recent NR simulations (available at Nedora \textit{et al.} 2020 \cite{nedora_vsevolod_2020_4283517}) with varying levels of realism in their treatment of the composition of the star and neutrino transport.  While they present fitting formulae with two distinct polynomial orders and including or excluding simulations with better/worse microphysics, that they then apply to a multitude of binary parameters, we only consider here their ``recommended'' fitting formula applied to the dynamical ejecta mass and disk mass.  The fitting formula is as follows:
\begin{equation}
    P_2^2(\tilde{Q},\tilde{\Lambda}) = b_0 + b_1 \tilde{Q} + b_2 \tilde{\Lambda} + b_3q^2 + b_4\tilde{Q}\tilde{\Lambda} + b_5\tilde{\Lambda}^2.
    \label{eq:NALejecta}
\end{equation}
For the dynamical ejecta, they provide best-fit parameters $b_0=-1.32$, $b_1=-0.382$, $b_2=-4.47\times10^{-3}$, $b_3=-0.339$, $b_4=3.21\times10^{-3}$, and $b_5=4.31\times10^{-7}$.  

\subsection{Disk Mass Models}

Fitting formulae for the mass remaining in an accretion disk around the remnant black hole (or neutron star) after a BNS merger have also been provided in multiple works. These formulae are typically fit to a lower number of simulations than dynamical ejecta formulae, as not all numerical relativity simulations report remnant disk masses. Additionally, the definition of the disk mass itself varies between different studies, complicating comparisons between different sets of simulations. Most notably, reported disk masses are sensitive to the time at which they are measured, and to the definition of the boundary between the ``disk'' and ``neutron star'' in neutron star-disk systems. They are however crucial to kilonovae modeling, as most BNS mergers likely eject more mass through disk winds than dynamical ejecta.

KF provides a fitting formula for the disk mass ejected from a binary neutron star merger, based on the subset of simulations used for the dynamical ejecta model that provide disk mass information. The disk mass fit for BNSs they arrive at is as follows: 
\begin{equation}
M_{disk}=M_1 \textrm{max} \{ 5\times 10^{-4},(aC_1+c)^d \}
\label{eq:KFdisk}
\end{equation}
with $C_1$ the compactness of the lower mass neutron star, and best-fit coefficients a=−8.1324, c=1.4820, and d=1.7784.  Their fit has an associated uncertainty of 
\begin{equation}
    \Delta M_{disk}=0.5M_{disk}+5 \times 10^{-4} M_\odot
    \label{eq:KFdeltadisk}
\end{equation}

We also consider the disk mass fitting formula provided by Dietrich \textit{et al.} 2020 ~\cite{Dietrich_2020}, hereafter DAL.  Their fit is derived from a series of 73 numerical relativity simulations performed by Kiuchi \textit{et al.} ~\cite{Kiuchi_2019}, Radice \textit{et al.} ~\cite{Radice_2018}, Dietrich \textit{et al.} ~\cite{Dietrich_2017b}, and Hotokezaka \textit{et al.} ~\cite{Hotokezaka_2011}.  Their model takes a similar form to their previous results in Coughlin \textit{et al.} 2019 ~\cite{Coughlin_2019}, but improves upon it by including mass-ratio dependent fitting parameters.  They arrive at the following disk mass fit:
\begin{multline}
    log_{10}(M_{disk})=max \left( -3, a \right.\\ \left. \left( 1 + \beta \textrm{tanh} \left( \frac{c-M/M_{threshold}}{d} \right) \right) \right) 
    \label{eq:DALdisk}
\end{multline} 
where $a$ and $b$ are given as
\begin{equation*}
    a= a_0+\delta a \xi 
\end{equation*}
\begin{equation*}
    b= b_0+\delta b \xi
\end{equation*}
and $\xi$ can be expressed as
\begin{equation}
    \xi = \frac{1}{2}\textrm{tanh} \left( \beta (Q-Q_{trans}) \right).
\end{equation}
Here, the best-fit coefficients are $a_0$=-1.581, $\delta a$=-2.439, $b_0$=-0.538, $\delta b$=-0.406, c=0.953, d=0.0417, $\beta$=3.910, and $Q_{trans}$=0.900.

Additionally, we consider the model provided by NAL.  Their polynomial fit for the dynamical ejecta (Equation ~\ref{eq:NALejecta}) is also used for the disk mass.  For the disk mass, they have that $b_0=-1.85$, $b_1=2.59$, $b_2=7.07\times10^{-4}$, $b_3=-0.733$, $b_4=-8.08\times10^{-4}$, and $b_5=2.75\times10^{-7}$.

Lastly, we employ the results from Radice \textit{et al.} 2018 ~\cite{Radice_2018}, hereafter RAL.  They study the mass ejection, nucleosynthetic yields, and electromagnetic counterparts of BNS mergers based on 59 high-resolution numerical relativity simulations.  They employ four nuclear-theory based equations of state (SFHo, BHB$\Lambda \phi$, DD2, and LS220) to constrain current estimates on merger outflows and ejecta properties.  Their models are calibrated to a mass ratio range of 0.85-1.0.  Their data is fit by the following expression:
\begin{equation}
    \frac{M_{disk}}{M_\odot}=max \left\{ 10^{-3}, \alpha + \beta \textrm{tanh} \left( \frac{\tilde\Lambda - \gamma}{\delta}\right) \right\}
\end{equation}
They also quote the uncertainty in the disk mass as 
\begin{equation}
    \Delta M_{disk}=0.5M_{disk}+5 \times 10^{-4} M_\odot
\end{equation}

\section{Methods}
\label{sec:methods}

\subsection{Comparing Dynamical Ejecta Models}
First, we compare dynamical ejecta models from DU, KF, and NAL.  We do this to see for what radii and mass ratios the models agree, and to measure their level of agreement over the parameter space.  For each figure, we fix the chirp mass, a quantity that is in practice obtained from GW observations.  We then generate a range of 100 mass ratios $Q=M_1/M_2$ from 0.5 to 1.0 (where $M_1<M_2$), which allows us to calculate the individual masses $M_1$ and $M_2$ with the following relation:
\begin{equation}
    M_2=M_c\frac{(Q+1)^{1/5}}{Q^{3/5}}.
\end{equation}
In practice, $M_c$ will typically be fairly well measured, but $Q$ will come with significant uncertainties. Additionally, measurements of $Q$ are partially degenerate with measurements of the spins of the neutron stars. In this manuscript, we focus on uncertainties due solely to the choice of the ejecta model, and thus assume fixed values of $M_c$ and $Q$; yet in actual observations the uncertainty in $Q$ would certainly have to be taken into account.

We then generate 100 values of the neutron star radius from 10km to 14km, working under the assumption that both neutron stars have the same radius; $R_1=R_2$.  The use of a constant radius is clearly an approximation. As evidenced by Figures ~\ref{fig:eosfigs_q9} and ~\ref{fig:eosfigs_q7}, however, a broad range of masses correspond to a very small range of radii for most equations of state consistent with existing nuclear physics and astrophysical constraints -- and visualizing models in the 2D space of mass ratios and radii is easier than in the infinite dimensional space of mass ratio and equations of state. We will lift this simplifying assumption later when considering predictions for specific equations of state.  

\begin{figure*}
    \centering
    \includegraphics[width=0.8\textwidth]{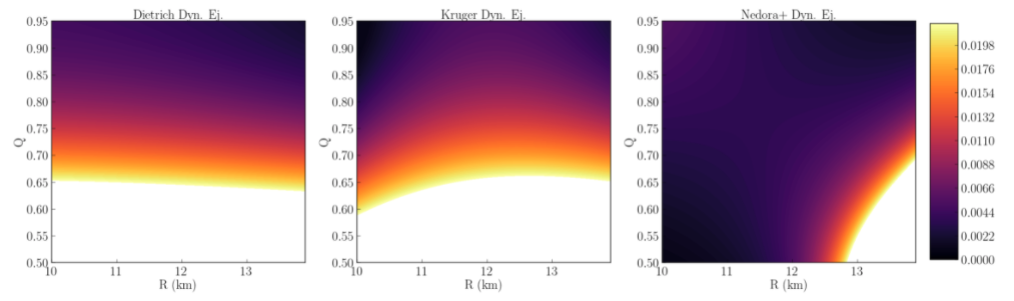} \\
    \includegraphics[width=0.8\textwidth]{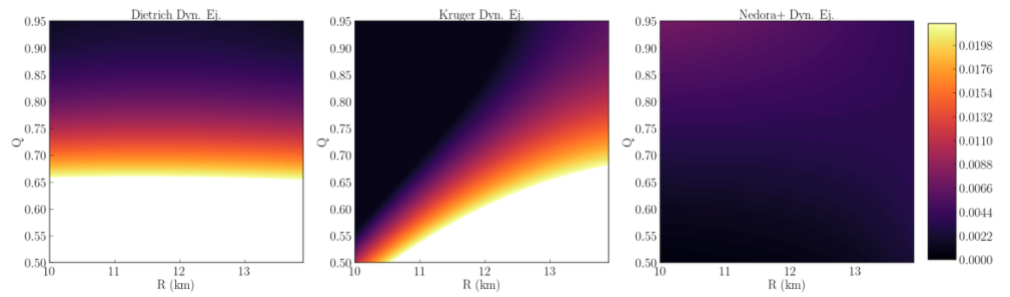} \\
    \caption{Dynamical ejecta mass predicted by Dietrich and Ujevic's model (left); Kruger and Foucart's model (center); and Nedora \textit{et al.}'s model (right).  The contours are in units of solar masses; the chirp mass is $M_c=1.186M_\odot$ for the top panel and $M_c=1.44M_\odot$ for the lower panel.  We truncate values above $0.022M_{\odot}$, to make the more reliable regions of the parameter space easier to study.  At high mass ratio, the extrapolation of fitting formulae outside of their range of applicability leads to extremely large predicted masses that have never been observed so far in simulations $\gtrsim 0.06M_\odot$).}
    \label{fig:dyn_ejecta_models}
\end{figure*}

We select chirp masses $M_c=1.44M_\odot$, the value associated with GW190425 ~\cite{Abbott_2020}, and $M_c=1.186M_\odot$, the value associated with GW170817 ~\cite{Abbott_2017}.  After calculating the component masses, the dynamical ejecta mass for each combination of R and Q is calculated according to Eqns. ~\ref{eq:DUdynej}, ~\ref{eq:KFdynej}, and ~\ref{eq:NALejecta}.  We then plot the dynamical ejecta from each model as color gradients on a radius vs. mass ratio grid, as seen in Figure ~\ref{fig:dyn_ejecta_models}.  

It is evident from Figure~\ref{fig:dyn_ejecta_models} that for both the DU and KF dynamical ejecta models, a higher ejecta mass is predicted for more unequal mass mergers.  This is to be expected, as unequal mass mergers are associated with more tidal deformation.  Additionally, KF's model is more sensitive to radius for both cases.  One objective of their model was to account for the expected lack of ejecta from highly compact stars near the equal mass limit, a phenomena that is physically understood but not automatically reflected by fitting formulae due to the lack of numerical simulations in that regime.  This is particularly visible in the higher chirp mass case, where a larger fraction of the systems are expected to collapse to a black hole. The NAL model shows very different qualitative behavior, especially outside of the regions where most numerical relativity simulations used to calibrate these models are found (near equal mass, non-collapsing systems).  We believe part of this discrepancy can be attributed to the fact that NAL's quadratic ejecta model takes a different functional form than KF and DU.  Additionally, it is a function of $\tilde{Q}$ and $\tilde{\Lambda}$, whereas the other two models are functions of the component masses and compactnesses.  It is important to note that while the KF model does recover what we expect to be the correct limit for small radii, there is no evidence that it performs better than the other models at more moderate radii. For example, Nedora \textit{et al.} 2021 showed that the NAL and KF models have, despite their very visible qualitative differences, comparable fitting residuals when fitted to their chosen datasets, while Camilletti {\it et al.} 2022 ~\cite{Camilletti:2022jms} showed that existing dynamical ejecta models compare poorly to numerical simulations for GW190425-like system (with the KF model performing especially poorly in that regime). There is also no particular reason to believe that one model is more accurate when extrapolating to very asymmetric systems -- in fact, all models predict unrealistically large ejecta mass for asymmetric systems, and it is more likely than not that they are all very inaccurate in that regime (the models predict masses well above $0.06M_\odot$, while no simulation used to calibrate the model has seen ejected masses above $0.06M_\odot$, and most simulations find dynamical ejecta masses $M_{\rm ej} \lesssim 0.01M_\odot$). Using these models for asymmetric binaries can thus be particularly dangerous if no corrections are applied, especially when assessing the impact of mergers on r-process nucleosynthesis (see e.g.~\cite{Chen:2021fro}).

We note that while a mass ratio $Q\sim 0.5$ is not physical for the low chirp mass case (the lower mass star would have $M\sim M_\odot$), it is a perfectly valid possibility for the high chirp mass case (lower mass of $\sim 1.2M_\odot$).

\subsection{Disk Mass Model Comparison}

Next, we perform the same procedure with the aforementioned disk mass models: KF (from ~\cite{Kr_ger_2020}), DAL (from ~\cite{Dietrich_2020}), RAL (from ~\cite{Radice_2018}), and NAL (from ~\cite{Nedora_2021}).  This was done to determine if any constraints about a BNS EM observation can be made given some level of agreement between said models.  

\begin{figure*}
    \centering
    \includegraphics[width=0.9\textwidth]{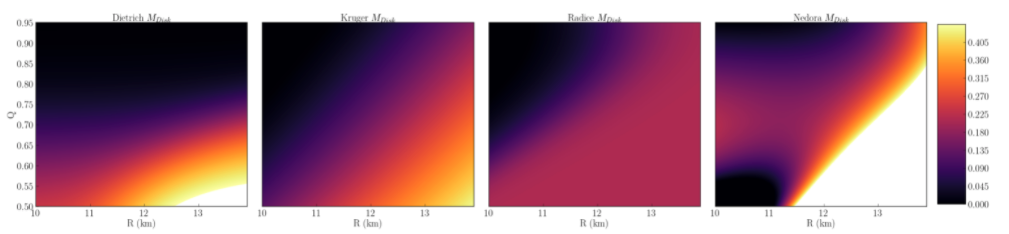} \\
    \includegraphics[width=0.9\textwidth]{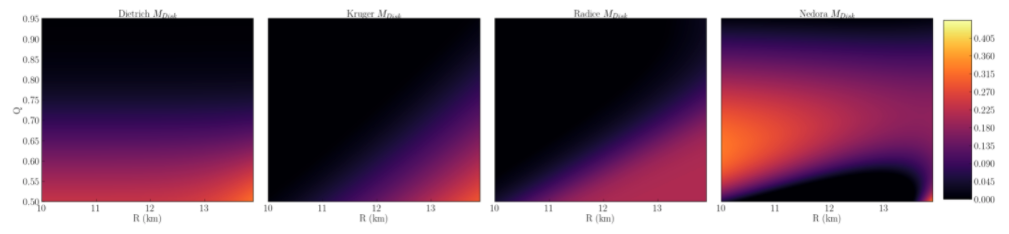} \\
    \caption{Disk mass ejecta calculated with each of the four disk mass models considered in this study.  The figures in the top panel have a chirp mass of $M_c=1.188M_\odot$; the figures in the bottom panel have $M_c=1.448M_\odot$.  We truncate the ejecta values for $M_{disk}>0.499M_{\odot}$ because the values quickly become unphysical outside of their domains.}
    \label{fig:disk_mass_models}
\end{figure*}

We can immediately observe that some predicted disk masses become unphysical for increasingly unequal mass cases, similar to the dynamical ejecta.  For example, the NAL model predicts upwards of 16$M_\odot$ of disk ejecta for low compactness (larger R, smaller Q) when $M_c=1.188M_\odot$.  For ease of visualization of the results, we truncate any values in excess of 0.5$M_\odot$ in our figures.  \\  All the disk mass models considered in this work predict little to no disk for nearly equal mass systems; this is to be expected, as the most efficient way to form massive accretion disks in BNS mergers is through the tidal disruption of a lower mass star by its more massive companion.  That less compact (larger R) stars produce more massive disks is well-captured in the DAL, KF, and RAL models.  NAL has a more complex behavior as a function of $R$, especially for high mass systems, due to the chosen functional form for the model (nearly all models from NAL are quadratic with a saddle point close to or within the fitting region).  

While BNSs with $Q\sim0.5$ are highly unlikely given the small range of possible NS masses, we deliberately probe a large parameter space to ascertain the models' behavior at extreme -- and less well-studied -- regions of the parameter space.

As for the dynamical ejecta models, the disk mass models show particularly significant disagreement for GW190425-like systems. Camilletti {\it et al.}~\cite{Camilletti:2022jms} showed that for GW190425-like systems, the KF model is accurate (within its relatively large error bars) in the parts of the high-mass parameter space covered by their simulations (for $Q\gtrsim 0.65$), though we caution again about inferring too much from that result about the quality of the model in other regions of the BNS parameter space.

\subsection{Equation of State}

The previous sections assumed constant-radius equations of state; an assumption convenient for visualizing the results but that does not exactly match realistic equations of state. For a more realistic treatment of the nuclear equation of state, we consider a suite of 2396 neutron star equations of state (EOSs).  The EOSs employed in this study are a series of best-fit ``spectral'' equations of state calibrated from GW170817 data.  More information about the EOSs included in this study can be found in Abbott \textit{et al.} 2018 ~\cite{Abbott_2018}, 
while the theoretical framework on which these equations of state are based can be found in Lindblom 2010
~\cite{Lindblom:2010bb}.  
Each EOS in the publicly available master file contains best-fit values for the spectral EOS parameters $\gamma_0$, $\gamma_1$, $\gamma_2$, and $\gamma_3$, as well as the pressure at a reference density.  Information about the mass, radius, and tidal deformability for each EOS can then be obtained from this data (see Raaijmakers \textit{et al.} 2020 ~\cite{raaijmakers2021challenges}).  \\
The main practical use of the analytical models discussed here is to attempt to determine whether the binary parameters of a given system - or a series of binary parameters from multiple events - can be used to reliably rule out certain equations of state, and thus constrain the true neutron star EOS.  In this manuscript, we test how robust these inferences are to the choice of model assuming extremely simplified constraints from potential observations. 

First, we define a fiducial system's ejecta mass, chirp mass and mass ratio.  In practice, the chirp mass and mass ratio are only known up to the potentially large uncertainties in gravitational wave parameter estimation, while the derivation of an ejecta mass from e.g. a kilonova signal is a non-trivial process that depends on the morphology, composition, and velocity of the ejecta as well as nuclear physics and radiation transport uncertainties. In order to directly study uncertainties due to the current modeling of the ejecta mass, however, we ignore these important complications for now, and do not consider any uncertainty in the fiducial ejecta mass, chirp mass, or mass ratio.  We define the fiducial ejecta mass as the total amount of dynamical ejecta and unbound disk ejecta; in practice, this quantity would have to be inferred from kilonova observations (see e.g. Raaijmakers \textit{et al.} 2021 ~\cite{raaijmakers2021challenges} and Coughlin \textit{et al.} 2019 ~\cite{Coughlin_2019}).  Defining $M_c$ and $Q$ sets the component masses of our system $M_1$ and $M_2$.  In the plots discussed in this section, $M_c$ and $Q$ are fixed in each subplot, and thus each subplot corresponds to some constant $M_1$ and $M_2$. 

Second, we obtain the radii $R_1$ and $R_2$ and tidal deformabilities $\Lambda_1$ and $\Lambda_2$ corresponding to $M_1$ and $M_2$ respectively, for each equation of state.  From these values, we can calculate the compactness $C$ of each neutron star as well as the reduced tidal deformability $\Tilde{\Lambda}$ of the binary.  These quantities all serve as inputs for the ejecta models considered in this work.

Then we calculate the dynamical ejecta as a function of the aforementioned inputs for each equation of state.  Each subplot considers \textit{either} the model presented by KF, DU, \textit{or} NAL; the dynamical ejecta model considered for each subplot is indicated in its title.  

Next, we calculate the disk ejecta, also functions of the aforementioned quantities, for each equation of state, assuming that a fraction $f_{\rm disk}$ of the accretion disk is ejected as disk wind in the seconds following the merger.  Given uncertainties $\sigma_{dyn}$ and $\sigma_{disk}$ in $M_{dyn}$ and $M_{disk}$ and assuming $f_{\rm disk}\in [f_{\rm low},f_{\rm high}]$, we calculate a ``window" of possible values for the total ejected mass:
    \begin{equation*}
        M_{min} = (M_{dyn}-\sigma_{dyn})+\textrm{f}_{low}(M_{disk}-\sigma_{disk})
    \end{equation*}
\begin{equation*}
        M_{max} = (M_{dyn}+\sigma_{dyn})+\textrm{f}_{high}(M_{disk}+\sigma_{disk})
\end{equation*}

Doing so provides a range of ejecta values that, according to the model, is compatible with the input parameters $M_c$, $Q$ and the chosen equation of state.  The fiducial ejecta mass is then compared to each calculated ejecta mass window to see if the fiducial mass falls within the calculated range.  For the KF and DU figures, we present a comparison of KF, DAL, and RAL's disk ejecta models.  For the NAL figures, we overlay the performance results of KF, DAL, RAL, and NAL's disk ejecta models.  This is done to more clearly draw a comparison between what ejecta models of different functional forms predict.  Additionally, we hold fixed the fractional range of unbound disk mass, with $f_{low}=0.1$ and $f_{high}=0.4$.  We initially considered two windows, $10\%-40\%$ and $20\%-25\%$, to check whether better constraints on the fraction of unbound disk mass impacts our results, but found that doing so does not provide any additional information at this point -- the differences between models and fit uncertainties have a stronger impact on our results. 

\begin{figure*}
\centering
    \includegraphics[width=6in]{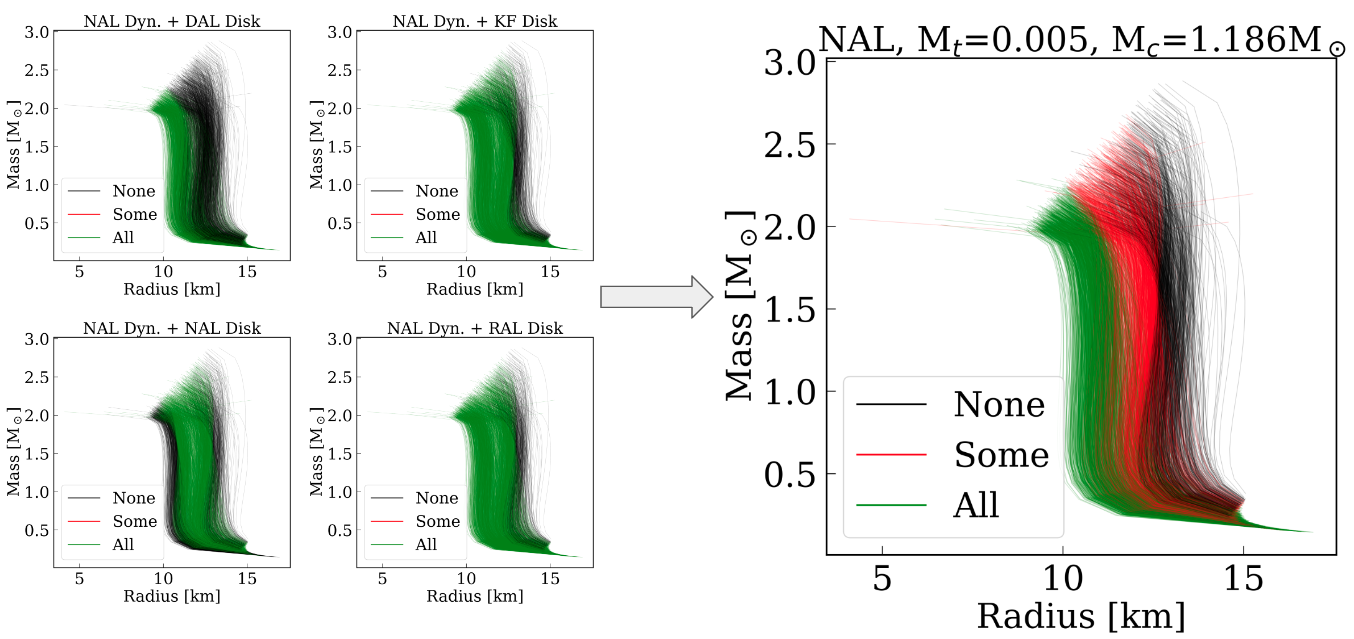} \\ % Made on July 15
    
\caption{A demonstration for how the subsequent figures in this paper are produced.  For the suite of EOSs considered in this work, we determine whether each one can reproduce ejecta in agreement with a pre-selected fiducial value, and color-code their mass-radius relations accordingly.  This is done by, for a given set of input parameters ($M_c$ and $Q$), calculating the dynamical ejecta according to one model, then calculating the disk ejecta according to multiple models for comparison, then overlaying where the results agree and disagree.  The larger figure shows an overlay of the results from (NAL dyn. ej.) + (disk ej.) for the four models considered in this work.  The four smaller subplots show the results for each of the disk models.} 
\label{fig:eos_demo}
\end{figure*}

If for a given equation of state the fiducial ejecta mass falls within the window of acceptable masses for all models under consideration, then that equation of state is assumed to be compatible with the fiducial observation. If it falls outside of that window for all models, the equation of state is incompatible with that observation. Finally, if the fiducial ejecta mass falls within the allowed region for some models but not other, no robust inference can be made about that equation of state -- and inferences made without considering modeling uncertainties may be inaccurate.  For each subplot, we overlay the mass-radius curve of each of the 2396 equations of state onto one plot, color coded according to whether the fiducial binary parameters agree with none (black), some (red), or all (green) of the disk mass ejecta models.  We demonstrate this process in Figure ~\ref{fig:eos_demo}.

\begin{figure*}
\centering

\includegraphics[width=7in]{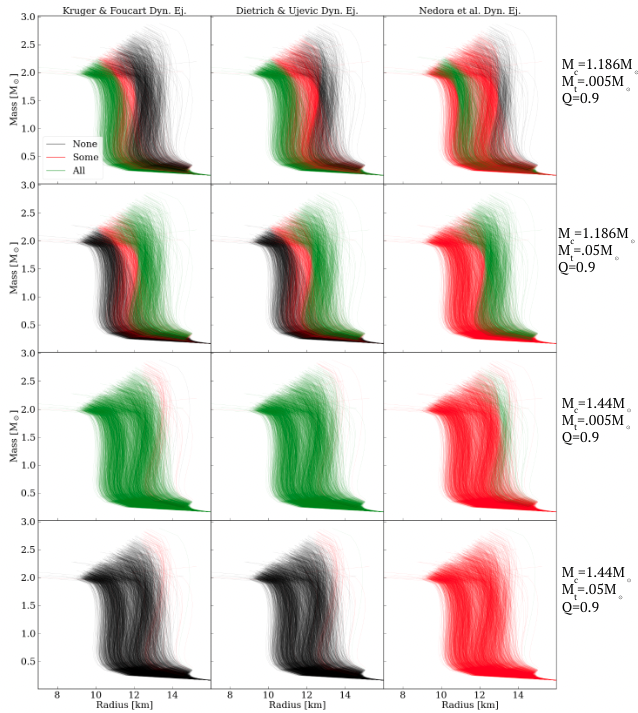} \\
\caption{Mass-radius relationships for 2396 equations of state from ~\cite{Abbott_2018}.  The figures above employ one of three dynamical ejecta models, KF, DU, or NAL; one of two fiducial ejecta masses, 0.05 and 0.005; and one of two chirp masses, $1.186M_\odot$ (in agreement with observations of GW170817) and $1.44M_\odot$ (in agreement with GW190425).  The KF and DU dyn. ej. figures overlay a comparison of KF, DAL, and RAL disk ejecta, whereas the NAL dyn. ej. figures overlay a comparison of KF, DAL, RAL, and NAL disk ejecta.  All figures here correspond to systems with a fixed mass ratio Q=0.9.}
\label{fig:eosfigs_q9}
\end{figure*}

\begin{figure*}
\centering

\includegraphics[width=7in]{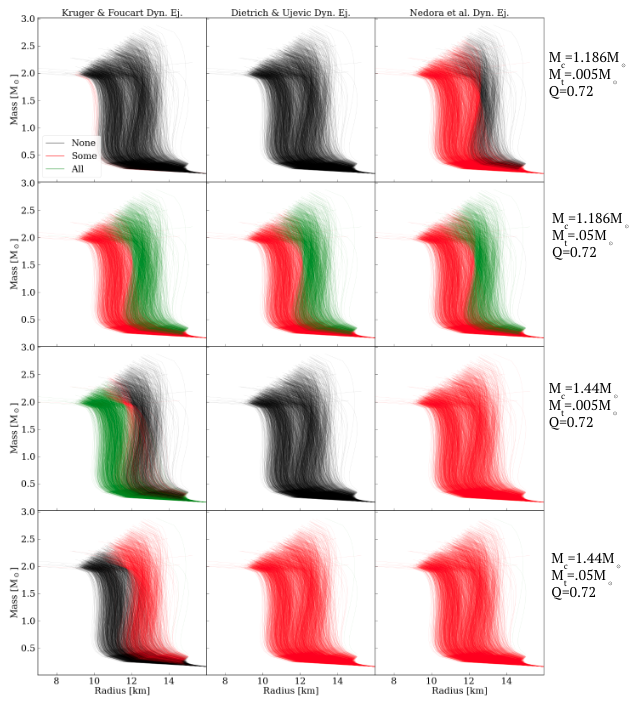} \\
\caption{The same scenarios as Figure \ref{fig:eosfigs_q9}, but with mass ratio $Q=0.72$.}
\label{fig:eosfigs_q7}
\end{figure*}

Our results are presented in Figures ~\ref{fig:eosfigs_q9} and ~\ref{fig:eosfigs_q7}.  We consider the same chirp masses as before, $1.186M_\odot$ and $1.44M_\odot$, and the two mass ratios $Q=0.9$ and $Q=0.72$.  We use $Q=0.72$ to demonstrate highly unequal mass mergers; any mass ratio more extreme produces values of $M_1$ and $M_2$ which fall outside the range allowed by some of the EOSs.  We ignore mass ratios $Q\lesssim0.7$ as we have already seen that the existing models are unreliable in that regime.  Finally, we initialize fiducial ejecta masses of $0.05M_\odot$, which is similar to GW170817, and $0.005M_\odot$, to study which systems are compatible with effectively negligible mass ejection (such as, possibly, GW190425).


We present all possible configurations of the input parameters in Figures ~\ref{fig:eosfigs_q9} and ~\ref{fig:eosfigs_q7}, with each row corresponding to the same $M_c$, $M_t$ and $Q$ for ease of comparison.  Looking first at Figure ~\ref{fig:eosfigs_q9}, we can see that for $M_c=1.448M_\odot$, the KF and DU models consistently agree that for all but the stiffest equations of state, it is not feasible for such a system to produce significant ejecta.  For such a system, NAL predicts the opposite: significant ejecta for all except the stiffest EOSs.  As can be seen in Figure ~\ref{fig:disk_mass_models}, DAL, KF, and RAL predict almost no ejecta to be produced for mass ratios above $Q \sim 0.75$, while NAL's disk model predicts the presence of disk ejecta up to around $Q \sim 0.9$.  This explains why the NAL figures appear red for columns 3 and 4 of Figure ~\ref{fig:eosfigs_q9}, while KF and DU are both green (for $M_t=0.005M_\odot$) and both black (for $M_t=0.05M_\odot$); NAL's disk model is in direct disagreement here with the other models.  

However, the models tend to agree in other parts of the parameter space.  This is shown in the second row of Figure ~\ref{fig:eosfigs_q9}, which is also the case most similar to GW170817 ($M_c=1.186M_\odot$, significant ejecta).  Here, only the least compact EOSs are consistently allowed.  We do observe some disagreement for the less massive EOSs; the KF and DAL subplots are able to consistently rule out approximately $30\%$ of the least stiff ones, but this not possible for the NAL figure.  This is likely because NAL's disk mass model produces relatively more ejecta for NSs with smaller radii compared to KF and DU.

We see that the trend for $M_c=1.186, Q=0.9, M_t=0.05M_\odot$ is inverted when we consider instead $M_t=0.005M_\odot$ of ejecta.  Here, only the most compact EOSs are able to predict such little mass.  Each subplot disallows a varying amount of EOSs: KF, DAL, and NAL each reject approximately $35\%$, $25\%$, and $7\%$ of the least compact EOSs.  Because each subplot shows this least compact $\sim 7\%$ as being disallowed, we can say that these EOSs would be reliably ruled out by such an observation.  

We now turn the focus of our discussion to Figure ~\ref{fig:eosfigs_q7}, which corresponds to $Q=0.72$.  Looking first to the top row which corresponds to $M_c=1.186M_\odot$ and a fiducial mass $M_t=0.005M_\odot$, we can see that KF and DAL subplots agree that such an unequal system cannot produce negligible ejecta.  The narrative shifts for the NAL subplot, where NAL's low dynamical ejecta appears to be the source of disagreement.  We can see from Figure ~\ref{fig:dyn_ejecta_models} that while DAL and KF predict significant dynamical ejecta for nearly all radii in this regime, NAL predicts almost none.  
The second row of Figure ~\ref{fig:eosfigs_q7}, which corresponds also to $M_c=1.186M_\odot$ but now a fiducial mass $M_t=0.05M_\odot$, demonstrates surprisingly consistent results.  (A close inspection shows that the NAL subplot has slightly less green-colored EOSs; the margin at which all models begin to agree occurs slightly to the right of where the margin exists for KF and DU.)  While this case does not robustly rule out certain EOSs, it does consistently identify viable ones -- here, the least compact EOSs are allowed.  
The fourth row of Figure ~\ref{fig:eosfigs_q7}, with $M_c=1.44M_\odot$ and $M_t=0.05M_\odot$ of ejecta, shows significant disagreement between ejecta models.  While the KF subplot disallows the most compact EOSs, DU and NAL do not arrive at this result.  That each subplot has around half the EOSs colored red in the same region is not informative; if anything, this demonstrates the models are consistently unable to make robust predictions in this region of parameter space.  
Similarly, we would be unable to draw robust conclusions from an observed system with $M_c=1.44M_\odot$ and $M_t=0.005M_\odot$.  When using DAL's dynamical ejecta model, all EOSs are disallowed; when using KF's dynamical ejecta model, we see the most compact EOSs allowed and least compact are consistently disallowed; and when using NAL's dynamical ejecta model, the models disagree for the entire parameter space.  While the results here disagree with each other, we can see that they are consistent with what is predicted in Figure ~\ref{fig:dyn_ejecta_models}.  DAL's dynamical ejecta model produces more mass at $Q=0.72$ for all radii, disqualifying all EOSs in the subplot.  We can also see from Figure ~\ref{fig:disk_mass_models} that NAL predicts, for most radii, a vastly larger disk than the other three disk models.

\section{Discussion}
We have considered a series of dynamical ejecta and disk mass models for binary neutron star mergers and estimated their predictive power by measuring agreement between models for a given system.  We find that regions where all existing models are consistent are quite rare, and that we should thus be cautious about making inferences about the parameters of BNS systems using just one of these models -- even when accounting for the models' stated uncertainties. We recommend to at least compare results with different ejecta models in order to get a sense of modeling uncertainties for a given system. 

In more detail, we find that for the dynamical ejecta models, DU and KF demonstrate a stronger dependence on mass ratio than is seen for NAL.  That more unequal mass systems produce relatively more ejecta compared to equal mass systems is well-captured in DU and KF's models, but not by NAL.  We note that this is likely explained by NAL's model being a function of different parameters compared to DU and KF ($Q$ and $\Tilde{\Lambda}$ vs. $M_{1,2}$ and $C_{1,2}$), and that the model takes a different functional form compared to the two other models.  Of the three models, KF demonstrates the strongest dependence on radius, which is an intentional feature meant to capture the different behavior of BNS systems of varying compactness.  All of the models begin to diverge to unrealistically high values as we probe increasingly unequal mass systems, which is due to the models being used outside of the parameter space to which they were calibrated.  For this reason, it can often be problematic to extrapolate ejecta fits.  We also wish to highlight that the differences between what the models predict is often of the order of the actual predicted ejecta mass.  

We perform the same study on the disk mass models considered in this work.  The models consistently predict low/no disk masses for compact (small R), equal-mass systems, which agrees with our current understanding that such mergers result in prompt collapse to a black hole.  However, different models have different thresholds for when ejecta is or is not produced.  All disk models with the exception of NAL produce significant ejecta for increasingly unequal mass systems; for $Q \sim 0.55-0.6$, the model predicts negligible disk mass.  As was the case for the dynamical ejecta, we see that the differences between the models are comparable in magnitude to the models' predictions.

Relative to NAL, we see stronger dependence on radius for the disk mass models from KF, DU, and RAL, which is particularly evident for the smaller chirp mass case of Fig. ~\ref{fig:disk_mass_models}.  For the high chirp mass panel of Fig. ~\ref{fig:disk_mass_models}, disagreement between models is most pronounced in the highly compact domain of the parameter space.  The lower chirp mass panel demonstrates divergence between models in several parts of the parameter space, notably in the top and bottom right corners.  As was the case for the dynamical ejecta, the differences between disk mass models is at times comparable in magnitude to the ejecta predicted by the models.  

We then studied a series of 2396 best-fit equations of state derived from aLIGO data for GW170817, and determined whether each equation of state could reproduce some injected fiducial ejecta mass given the dynamical ejecta and disk mass ejecta models we have considered.  We find that there is slightly stronger agreement between models for the Q=0.9 case, with some fiducial observations consistently ruling out/accepting a subset of the proposed equations of state. However, for most fiducial observations (and particularly for more asymmetric systems), the large variations in the predictions of different fitting formulae should urge us to proceed with extreme caution when attempting to derive information about the equation of state of neutron stars from kilonovae signals powered by BNS mergers. 

For the case most similar to GW170817 ($M_c=1.186M_\odot$, Q=0.9, $M_{tot}=0.05M_\odot$), we see all models agree that the least compact (largest radius and, typically, largest maximum mass) equations of state are consistent with such a system.  This is one of the only spaces where all three subfigures produce consistent results.  Most models predict that the more compact equations of state should be ruled out by such an observation (and NAL, which does not at $Q=0.9$, would rule out compact neutron stars as well if we had chosen a slightly more symmetric mass ratio).

Notably, we are unable to draw any conclusions about the NS EOS from the case most similar to GW190425 ($M_c=1.44M_\odot$, $Q=0.72$, negligible ejecta).  As evidenced in Figure~\ref{fig:eosfigs_q7}, each dynamical ejecta model provides a different result.  With KF's model, the most compact EOSs are allowed and the lesser are ruled out; DU's model disallows all EOSs; and we see that no claims can be made about the EOS when using NAL's model.  That the different models are able to produce such discrepant results for unequal mass systems suggests that further simulations in this regime are needed in order to properly calibrate future ejecta models to a broader region of parameter space.    
We should also note that this study is highly idealized, neglecting uncertainties due to the ejecta composition and geometry, nuclear reactions, and photon transport in the outflows. Our only objective is to assess (dis)agreement between existing outflow models, and we see that even if differences between models were the sole source of uncertainty, that uncertainty would already be extremely limiting.  However, with increasingly detailed simulations that cover a larger set of parameters, we hope that in the future these models can be used to place robust constraints on the amount of ejecta from NSNS mergers.

\section*{Acknowledgements}
A.H. and F.F. gratefully acknowledge support from the DOE through grant DE-SC0020435, and from NASA through grant 80NSSC18K0565.  GR and SMN are grateful for financial support from the Nederlandse Organisatie voor Wetenschappelijk Onderzoek (NWO) through the Projectruimte and VIDI grants (Nissanke).  SMN also acknowledges financial support from the NWO sector plan.

\bibliography{minibib.bib}

\end{document}